\documentclass[aps,prl,showpacs,floatfix,twocolumn,showpacs]{revtex4-1}
\usepackage{epsfig,graphicx}
\usepackage{bm}
\usepackage{color}

\usepackage[latin1]{inputenc}
\newcommand{\beq}{\begin{equation}}
\newcommand{\eeq}{\end{equation}}

\newcommand{\bea}{\begin{eqnarray}}
\newcommand{\eea}{\end{eqnarray}}

\newcommand{\REM}[1]{\textcolor{green}{}}

\usepackage{amssymb,amsfonts,amsmath}

\begin{document}
\title{Two-dimensional Turbulence in Symmetric Binary-Fluid Mixtures: Coarsening Arrest by the Inverse Cascade}
\author {Prasad Perlekar$^{1}$, Nairita Pal$^{2}$, and Rahul Pandit$^{2,3}$} 
\affiliation{${}^1$ TIFR Centre for Interdisciplinary Sciences, 21 Brundavan Colony, Narsingi, Hyderabad 500075, India \\
${}^2$ Centre for Condensed Matter Theory, Indian Institute of Science, Bangalore 560012, India \\
 ${}^3$ Jawaharlal Nehru Centre for Advanced Scientific Research, Jakkur, Bangalore, India}

\begin{abstract}
We study two-dimensional (2D) binary-fluid turbulence by carrying out an
extensive direct numerical simulation (DNS) of the forced, statistically steady
turbulence in the coupled Cahn-Hilliard and Navier-Stokes equations. In the
absence of any coupling, we choose parameters that lead (a) to spinodal
decomposition and domain growth, which is characterized by the spatiotemporal
evolution of the Cahn-Hilliard order parameter $\phi$, and (b) the formation of
an inverse-energy-cascade regime in the energy spectrum $E(k)$, in which energy
cascades towards wave numbers $k$ that are smaller than the energy-injection
scale $k_{inj}$ in the turbulent fluid. We show that the
Cahn-Hilliard-Navier-Stokes coupling leads to an arrest of phase separation at
a length scale $L_c$, which we evaluate from $S(k)$, the spectrum of the
fluctuations of $\phi$. We demonstrate that (a) $L_c \sim L_H$, the Hinze scale
that follows from balancing inertial and interfacial-tension forces, and (b)
$L_c$ is independent, within error bars, of the diffusivity $D$.  We elucidate
how this coupling modifies $E(k)$ by blocking the inverse energy cascade at a
wavenumber $k_c$, which we show is $\simeq 2\pi/L_c$. We compare our work with
earlier studies of this problem. 
\end{abstract}


\maketitle

Two-dimensional (2D) fluid turbulence, which is of central importance in a
variety of oceanographic and atmospheric flows, is fundamentally different from
three-dimensional (3D) fluid turbulence as noted in the pioneering studies of
Fj{\o}rtoft, Kraichnan, Leith, and Batchelor~\cite{fjo53,kra67,lei68,bat69,Les97}.  In
particular, the fluid-energy spectrum in 2D turbulence shows (a) a
\textit{forward cascade} of enstrophy (or the mean-square vorticity), from the
energy-injection wave number $k_{inj}$ to larger wave numbers, \textit{and} (b)
an \textit{inverse cascade} of energy to wave numbers smaller than $k_{inj}$.
We elucidate the arrest of phase separation in a 2D, symmetric, binary-fluid
mixture by turbulence.

In the absence of turbulence, binary-fluid mixtures have played a pivotal role
in the development of the understanding of (a) equilibrium critical phenomena
at the consolute point, above which the two fluids
mix~\cite{fish67,kum83,Kar07}, (b) of nucleation~\cite{hua74}, and (c)
spinodal decomposition, the process by which a binary-fluid mixture, below the
consolute point and below the spinodal curve, separates into the two,
constituent liquid phases until, in equilibrium, a single interface separates
the two coexisting phases~\cite{gun83,onu02}. In the late stages of
growth, as the binary-fluid mixture evolves via spinodal decomposition towards
the completely phase-separated, equilibrium state, the domains of these two
phases coarsen to yield ever larger domains whose linear size diverges as a
power of the time $t$; this divergence leads to universal scaling forms for the
time-dependent correlation functions~\cite{lif59,fur85,sig79,bra94,wag98,ken00,ken01,onu02,puri09,cat12,datt15} 
of the order parameter $\phi$, which distinguishes the two phases of the binary-fluid mixture.

Coarsening arrest by 2D turbulence has been studied in Ref.~\cite{ber05}, where
it has been shown that, for length scales smaller than the energy-injection
scale $\ell_{inj}=2\pi/k_{inj}$, the typical linear size of domains is
controlled by the average shear across the domain.  However, the nature of
coarsening arrest, for scales larger than $\ell_{inj}$, i.e., in the
inverse-cascade regime, still remains elusive. In particular, it is not clear
what happens to the inverse energy transfer, in a 2D binary-liquid, turbulent
mixture, in which the mean size of domains provides an additional, important
length scale. We resolve these two issues in our study. By combining
theoretical arguments with extensive direct numerical simulations (DNSs) we
show that the Hinze length scale $L_H$ (see Refs.~\cite{per14,hin55}) provides
a natural estimate for the arrest scale; and the inverse flux of energy also
stops at a wave-number scale $\simeq 2\pi/L_H$. In particular, we study
two-dimensional (2D) binary-fluid turbulence by carrying out a 
direct numerical simulation (DNS) of the forced, statistically steady
turbulence in the coupled Cahn-Hilliard and Navier-Stokes equations. In the
absence of any coupling, our choice of forcing leads (a) to spinodal decomposition
and domain growth, which we examine by the spatiotemporal evolution of $\phi$, and
(b) to the formation of an inverse-energy-cascade regime in the energy spectrum
$E(k)$, in which energy cascades towards wave numbers $k$ that are smaller than
the energy-injection scale $k_{inj}$ in the turbulent fluid. We show that the
Cahn-Hilliard-Navier-Stokes coupling leads to an arrest of phase separation at
a length scale $L_c$, which we evaluate from $S(k)$, the spectrum of the
fluctuations of $\phi$. We demonstrate (a) $L_c \sim L_H$ and (b) that $L_c$ is
independent, within error bars, of the diffusivity $D$.  We elucidate how this
coupling modifies $E(k)$ by blocking the inverse energy cascade at a wavenumber
$k_c$, which we show is $\simeq 2\pi/L_c$.

We model a symmetric binary-fluid mixture by using the
incompressible Navier-Stokes equations coupled to the
Cahn-Hilliard or Model-H equations~\cite{hoh77,cah68}. We are
interested in 2D incompressible fluids, so we use the following
stream-function-vorticity formulation~\cite{per09,per09b,bof12} 
for the momentum equation: 
\begin{eqnarray}
%\nonumber
(\partial_t + \bm u \cdot \nabla) \omega &=& \nu \nabla^2 \omega - \nabla \times (\phi \nabla \mu) + f_\omega,  \label{ch:eq1} \\
%\nonumber
(\partial_t + {\bm u} \cdot \nabla) {\phi} &=& M \nabla^2 {\mu},~{\rm{and}}~\nabla \cdot {\bm u} =0.
\label{ch:eq2}
\end{eqnarray}
Here ${\bm u}\equiv(u_x,u_y)$ is the fluid velocity, $\omega=(\nabla \times
{\bm u}) {\hat{\bm e}}_z$, $\phi({\bm x},t)\in[-1,1]$ is the Cahn-Hilliard
order parameter at the point ${\bm x}$ and time $t$,  $p$ is the pressure,
$\mu({\bm x},t)=\delta {\mathcal F}[\phi]/\delta \phi({\bm x},t)$ is the
chemical potential, ${\mathcal F}[\phi]=\Lambda \int [(\phi^2-1)^2/(4\xi^2) +
|\nabla \phi|^2/2] d{\bm x}$ is the free energy, $\Lambda$ is the mixing energy density, 
$\xi$ controls the width of the interface between the two phases of the binary-fluid mixture, $\nu$ is the
kinematic viscosity, the surface tension $\sigma=2\sqrt{2}/3 (\Lambda/\xi)$,
the mobility of the binary-fluid mixture is $M$, and $f_{\omega}$ is the
external driving force. For simplicity, we study mixtures in which  $M$ is
independent of $\phi$ and both components have the same density and
viscosity~\cite{ken01}. We use periodic boundary conditions in our square
simulation domain, with each side of length $L=2 \pi$. To obtain a substantial
inverse-cascade regime, we stir the fluid at an intermediate length scale by
forcing in Fourier space in a spherical shell with wave-number
$k_{inj}=2\pi/\ell_{inj}$. Our choice of forcing
$\hat{f}_\omega({\mathbf k},t)=\hat{\omega}({\mathbf k},t)/\sum_{k=k_{inj}} \hat{\omega}({\mathbf k},t)$,
where the caret indicates a spatial Fourier transform, ensures that there is a constant
enstrophy-injection rate. In all our studies we use $k_{inj}=40$ so that there
is a clear separation between $\ell_{inj}$ and $\xi$. We conduct DNSs of Eqs.~\eqref{ch:eq1} and \eqref{ch:eq2} by using
a pseudospectral method~\cite{Can88}; because of the cubic nonlinearity in the
chemical potential $\mu$, we use $N/2$-dealiasing. For time integration we use
the exponential Adams-Bashforth method ETD2~\cite{cox02}. Important
nondimensional numbers for the turbulent flows here are the Grashof number $Gr
\equiv (L^4f_\omega/\nu^2)$, the injection-scale Reynolds number $Re\equiv
u_{inj} \ell_{inj}/\nu$, with $u_{inj}=(\epsilon_{inj} \ell_{inj})^{1/3}$,
where $\epsilon_{inj}$ is the energy-injection rate, the Weber number $We\equiv
\rho u_{inj}^2 \ell_{inj}/\sigma$, the Cahn number $Ch = \xi/L$, the Peclet
number $Pe \equiv u_{rms}L/D$, where $u_{rms}$ is the root-mean square
velocity, and the Schmidt number $Sc \equiv \nu/D$, where $D\equiv M\Lambda/\xi^2$ 
is the diffusivity of our binary-fluid mixture.  We give the parameters 
for our simulations in the Supplemental Material~\cite{suppmat}.  

Given ${\bm u}({\bm x},t)$ and $\phi({\bm x},t)$ from our DNS, we calculate the
energy and order-parameter (or phase-field) spectra, which are, respectively,
$E(k) \equiv \sum_{k-\frac{1}{2} \leq k' \leq k+\frac{1}{2}} \langle
|\hat{\bm u}({\mathbf k}',t)|^{2} \rangle_{t}$ and $S(k) \equiv \sum_{k-\frac{1}{2}
\leq k' \leq k+\frac{1}{2}} \langle |\hat{\phi}({\mathbf k}',t)|^{2}
\rangle_{t}$, where $\langle \rangle_{t}$ denotes the average over time in
the statistically steady state of our system.  The total kinetic energy is
$E(t)=\frac{1}{2}\langle |{\bf u}({\bm x},t)|^{2} \rangle_{{\bm x}}$ and the
total enstrophy $\epsilon(t)=\frac{1}{2} \langle |\omega({\bf x},t)|^{2}
\rangle_{{\bf x}}$, where $\langle \rangle_{{\bm x}}$ denotes the average over
space, $\langle
f_{\omega} \omega \rangle$ is the enstrophy-injection rate, which is related to
the energy-injection rate via $\epsilon_{inj}=\langle f_{\omega} \omega
\rangle/k_{inj}^2$, $E=0.5 \sum_k E(k)$ is the fluid kinetic energy,
$\epsilon_\nu=\nu \sum_k k^2 E(k)$ is the fluid-energy dissipation rate, and 
$\epsilon_\mu= M \sum_k k^2 \langle |\hat{\mu}({\mathbf k},t)|^2| \rangle_t$  is the 
energy-dissipation rate because of the phase field $\phi$.

Forced, 2D, statistically steady, Navier-Stokes-fluid turbulence displays a
forward cascade of enstrophy, from $\ell_{inj}$ to smaller length scales, and
an inverse cascade of energy to length scales smaller than $\ell_{inj}$. In the
inverse-cascade regime, on which we concentrate here, $E(k) \sim k^{-5/3}$
(see, e.g., Refs.~\cite{kra67,Les97}) and the energy flux $\Pi(k) \sim
\epsilon \equiv \langle \epsilon(t) \rangle_t$ assumes a constant value. For
the Cahn-Hilliard model, if it is \textit{not} coupled to the Navier-Stokes
equation, $S(k,t) \sim {\mathcal S}(k \mathbb{L}(t))$, for large times, where
the time-dependent length scale $\mathbb{L}(t)\sim t^{1/3}$, in
the early Lifshitz-Slyozov~\cite{lif59,bra94,puri09,cat12} regime; if the
Cahn-Hilliard model is coupled to the Navier-Stokes equation, then, in the
absence of forcing, $\mathbb{L}(t)\sim t$, in the viscous-hydrodynamic regime,
first discussed by Siggia~\cite{sig79,bra94,puri09,cat12}, and
$\mathbb{L}(t)\sim t^{2/3}$, in the very-late-stages in the Furukawa~\cite{fur85} and
Kendon~\cite{ken00} regimes. For a discussion of these regimes and a
detailed exploration of a universal scaling form for $\mathbb{L}(t)$ in 3D we
refer the reader to Ref.~\cite{ken01}. We now elucidate how these scaling
forms for $E(k)$ and $S(k,t)$ are modified when we study forced 2D turbulence,
in the inverse-cascade regime in the coupled
Cahn-Hilliard-Navier-Stokes equations.

\begin{figure}[!h]
\begin{center}
\includegraphics[width=0.24\linewidth]{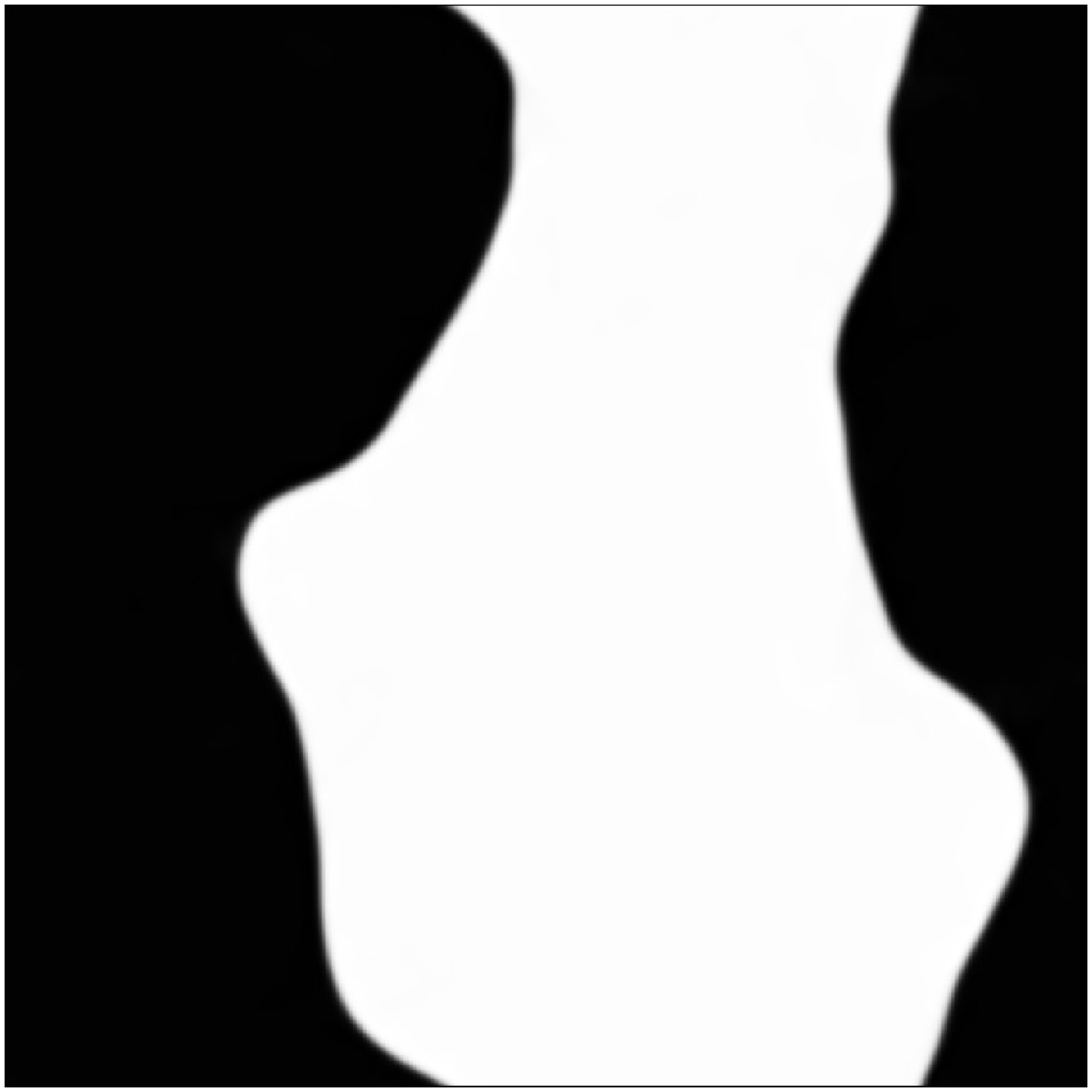}
\includegraphics[width=0.24\linewidth]{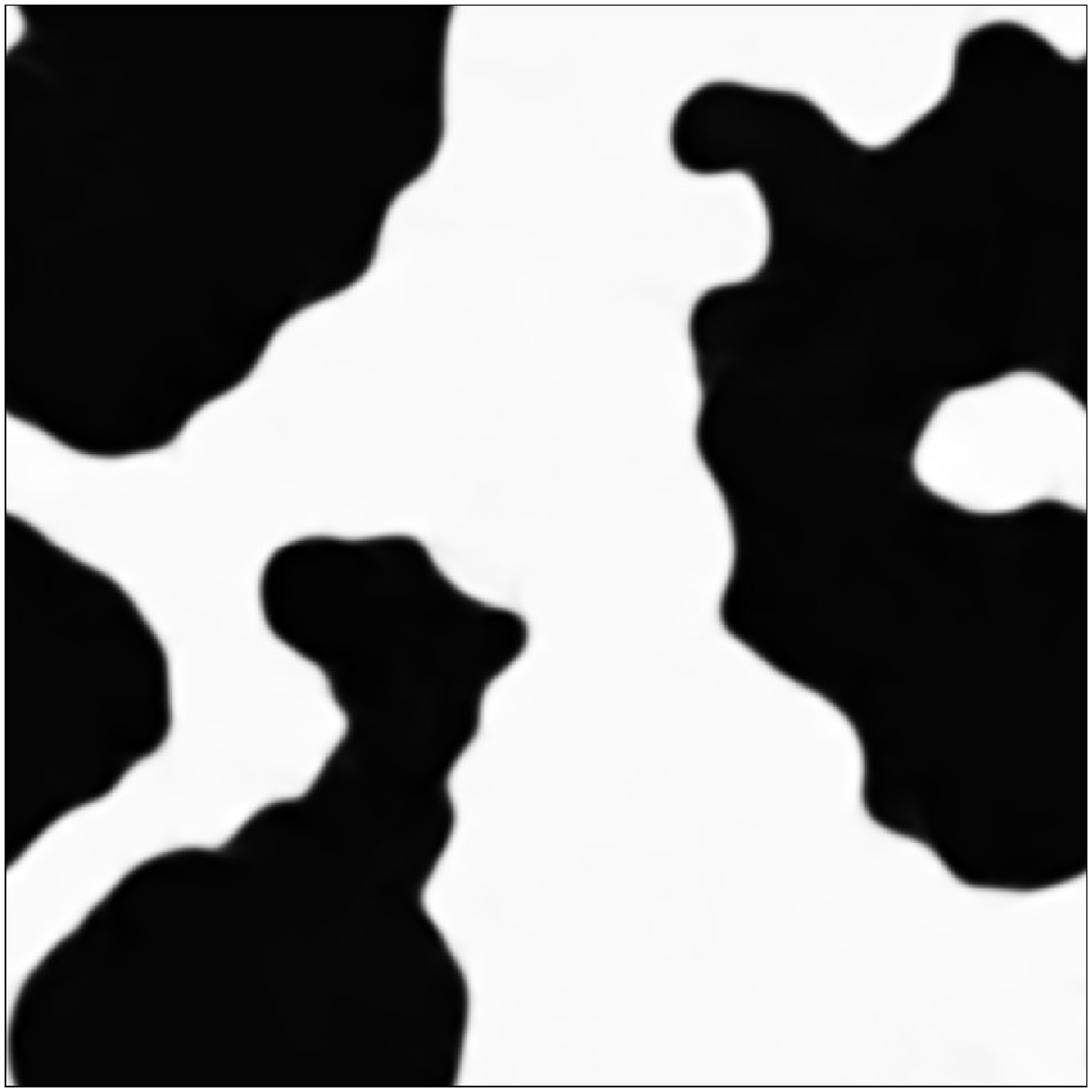}
\includegraphics[width=0.24\linewidth]{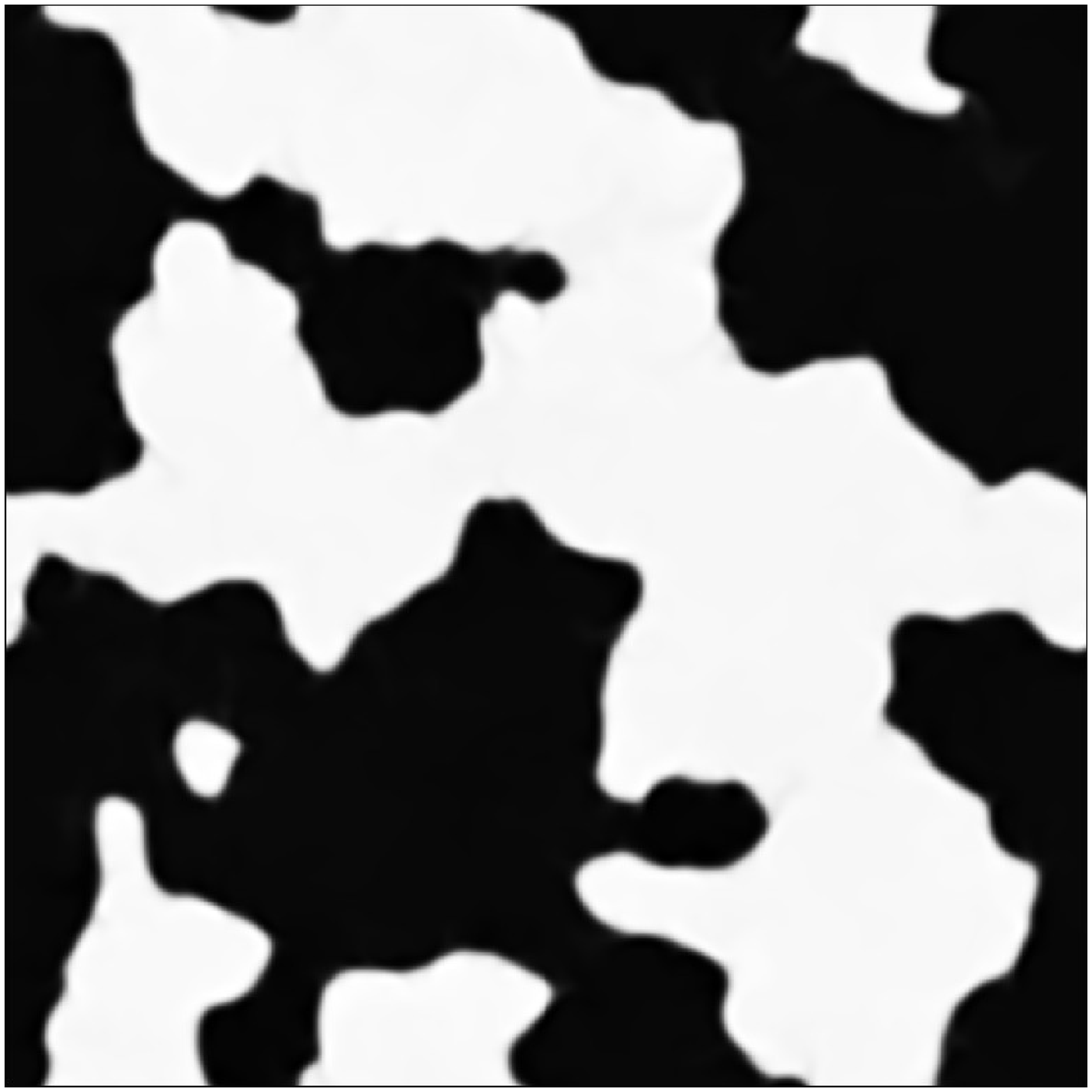}
\includegraphics[width=0.24\linewidth]{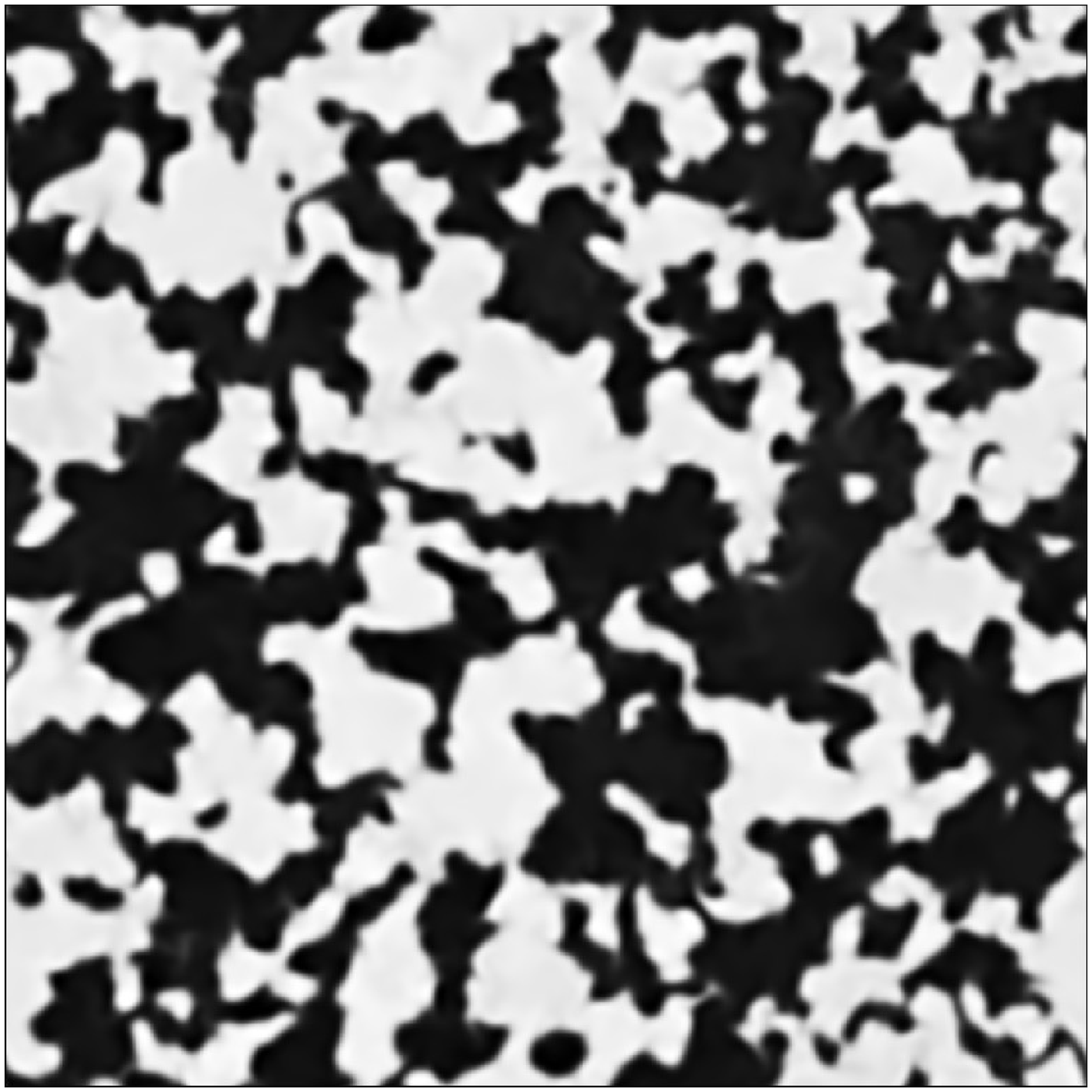}
\end{center}

\caption{\label{fig:fig1} Pseudo-gray-scale plots of the order parameter field
$\phi$, at late times when coarsening arrest has occured, in 2D
symmetric-binary-fluid turbulence with $Re=124$. Note that the domain size
decreases as we  increase the Weber number $We$ from the leftmost to the rightmost panel: $We=1.2\cdot 10^{-2}$ (${\tt R3}$);  
$We=5.9\cdot 10^{-2}$ (${\tt R4}$);  $We=1.2\cdot10^{-1}$ (${\tt R5}$); and $We=5.9\cdot 10^{-1}$ (${\tt R8}$).}
\end{figure}

In Fig.~\ref{fig:fig1} we show pseudo-gray-scale plots of $\phi$, at late times
when coarsening arrest has occurred, for four different values of $We$ at
$Re=124$; we find that the larger the value of $We$ the smaller is the linear
size that can be associated with domains; this size is determined by the
competition between turbulence-shear and interfacial-tension forces.  This
qualitative effect has also been observed in earlier studies of 2D and 3D
turbulence of symmetric binary-fluid mixtures \cite{cha87,rui81,pine84,aro84,has95,lac95,onu97,berth01,ber05}.  

\begin{figure*}
\includegraphics[width=0.32\linewidth]{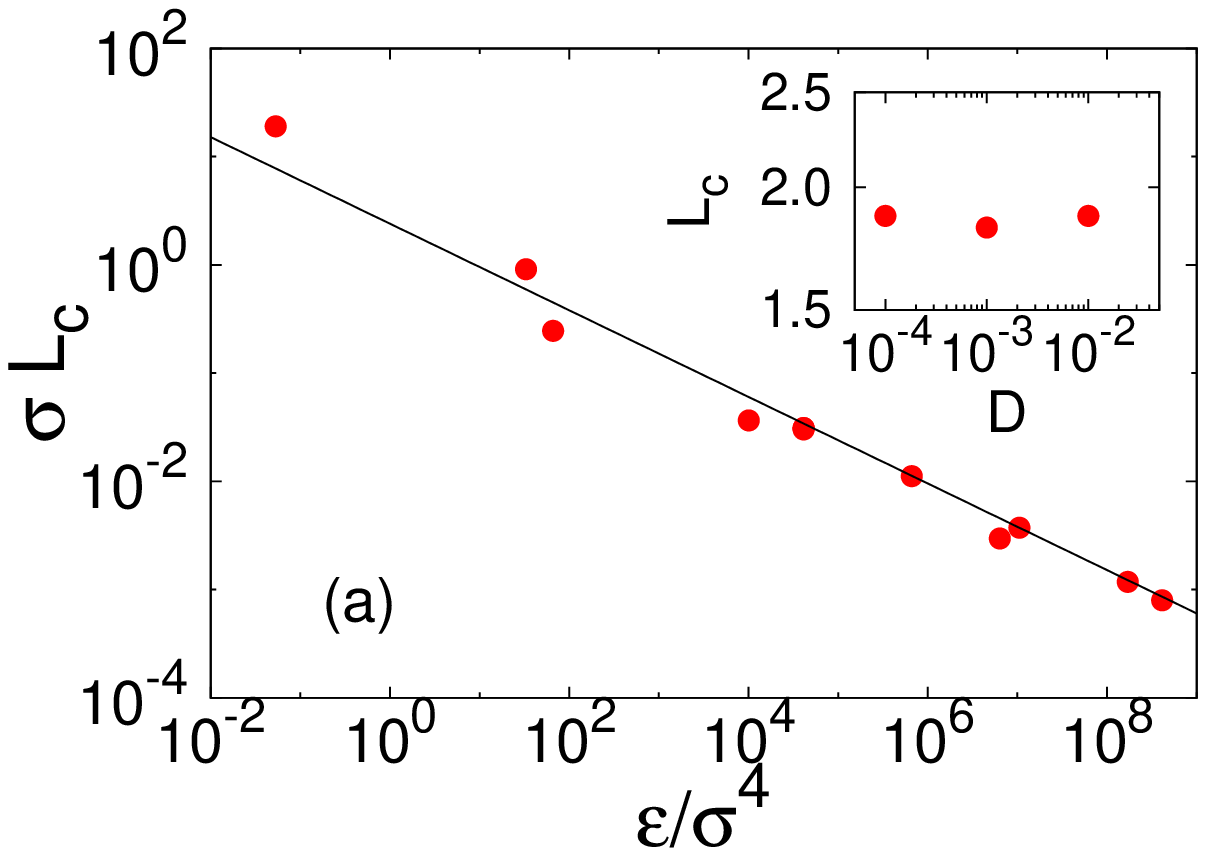}
\includegraphics[width=0.32\linewidth]{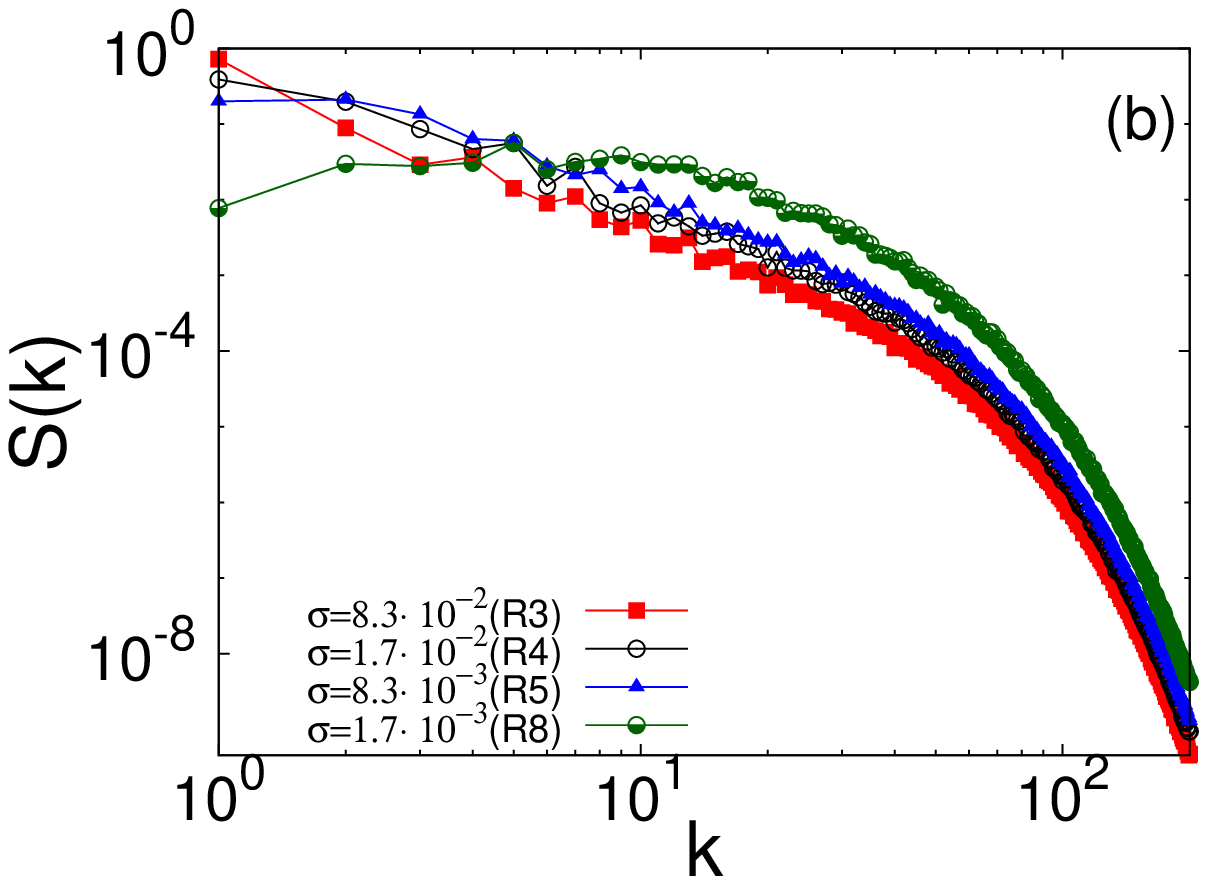}
\includegraphics[width=0.32\linewidth]{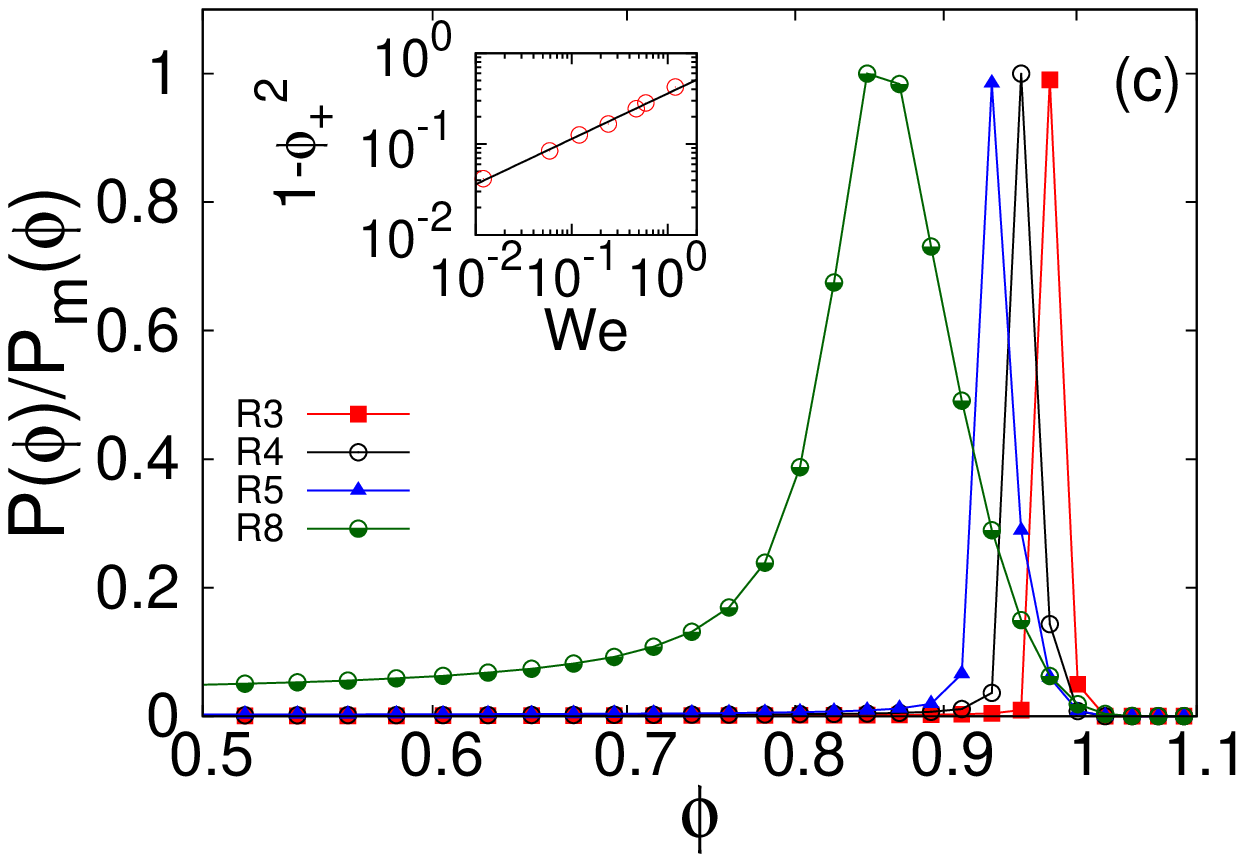}
\caption{\label{fig:figh}(Color online)(a) Log-log (base 10) plot of $\sigma L_c$ versus
$\epsilon/\sigma^4$ showing data points ($L_c$ from Eq.~\eqref{eq:lc}, with
$S(k)$ from our DNS) in red.  The black line is the Hinze result~(\ref{eq:eqh})
for $L_H$; a fit to our data yields a constant of proportionality $\simeq 1.6$
and an excellent approximation to the arrest scale $L_c$ over several orders of
magnitude on both vertical and horizontal axes; the plot of $L_c$ versus $D$,
in the inset, shows that, for fixed values of $\epsilon_\nu$ and $\sigma$ (runs
{\tt R1,R2} and {\tt R4}), $L_c$  is independent of $D$ (within error bars), as
is implied by the Hinze condition (see text). (b) Log-log (base 10) plots of the spectrum $S(k)$, of
the phase-field $\phi$, versus $k$; as $We$ increases (i.e., $\sigma$
decreases) the low-$k$ part of $S(k)$ decreases and $S(k)$ develops a broad and
gentle maximum whose peak moves out to large values of $k$. (c) Plots versus $\phi$,
in the vicinity of the maximum at $\phi_+$, of the normalized PDFs
$P(\phi)/P_m(\phi)$, where $P_m(\phi)$ is the maximum of $P(\phi)$; the peak
position $\phi_+ \to 1$ as $We$ increases (see the inset which suggests that
$1-\phi_+^2 \sim We^{1/2}$ (black line)). } 
\end{figure*}

We calculate the coarsening-arrest length scale  
\begin{equation}
L_c=2 \pi[\sum_k S(k)]/[\sum_k k S(k)].
\label{eq:lc}  
\end{equation}
We now show that $L_c$ is determined  by the Hinze scale
$L_H$, which we obtain, as in Hinze's pioneering study of
droplet break-up~\cite{hin55}, by balancing the surface tension with 
the inertia as follows:

\begin{equation}
L_H \sim \epsilon_{inj}^{-2/5}\sigma^{3/5}.
\label{eq:eqh}
\end{equation} 
We obtain for 2D, binary-fluid turbulence the intuitively appealing result $L_c
\sim L_H$ (for a similar, recent Lattice-Boltzmann study in 3D see 
Ref.~\cite{per14}). In particular, if we determine $L_c$
from Eq.~\eqref{eq:lc}, with $S(k)$ from our DNS, we obtain the red points in
Fig.~\ref{fig:figh}, which is a log-log plot of $\sigma L_c$ versus
$\epsilon_{inj}/\sigma^4$; the black line is the Hinze result~\eqref{eq:eqh} for $L_H$,
with a constant of proportionality that we find is $\simeq 1.6$ from a fit to our
data.  We see from Fig.~\ref{fig:figh} that the Hinze length scale $L_H$ gives
an excellent approximation to the arrest scale $L_c$ \textit{over several orders of
magnitude on both vertical and horizontal axes}.  Note that the Hinze estimate
also predicts that, for fixed values of $\epsilon_{inj}$ and $\sigma$, the 
coarsening-arrest scale is independent of $D$; the plot of $L_c$ versus $D$, in
the inset of Fig.~\ref{fig:figh}, shows that our data for $L_c$ are consistent
(within error bars) with this prediction.

In Fig.~\ref{fig:figh} (b) we show clearly how the arrest of coarsening
manifests itself as a suppression of $S(k)$, at small $k$ (large length
scales). This suppression increases as $We$ increases (i.e., $\sigma$
decreases); and $S(k)$ develops a broad and gentle maximum whose peak moves out
to large values of $k$ as $We$ grows. These changes in $S(k)$ are associated
with $We$-dependent modifications in the probability distribution function
(PDF) $P(\phi)$ of the order parameter $\phi$, which is symmetrical about $\phi
= 0$ and has two peaks at $\phi = \phi_{\pm}$, where $\phi_+ = -\phi_- > 0$; we
display $P(\phi)/P_m(\phi)$ in Fig.~\ref{fig:figh} (c) in the vicinity of  the peak at
$\phi_+$; as $We$ increases, $\phi_+$ decreases; here $P_m(\phi)$ is the maximum value
of $P(\phi)$. In particular, our DNS
suggests that $1-\phi_+^2 \sim We^{1/2}$, for small $We$.

The modification in $P(\phi)$ can be understood qualitatively by making the approximation that the effect of the fluid on the equation for $\phi$ can be encapsulated into an eddy diffusivity $D_e$~\cite{aro84, nar07}. 
The eddy-diffusivity-modified Cahn-Hilliard equation is $\partial_t \phi = (D_e-D) \nabla^2 \phi + D \nabla^2 \phi^3 + M \Lambda \nabla^4 \phi$, which gives the maximum 
and minimum values of $\phi$ as $\phi_\pm=\sqrt{(D-D_e)/D}$.  Furthermore, if we neglect the nonlinear term~\cite{bra94,cat12}, we find easily that the modified growth rate 
is $D k^2[(1-D_e/D)-M\Lambda k^2]$; i.e., all wave numbers larger than $k_d=\sqrt{(1-D_e/D)/(\Lambda M)}$ are stable to perturbations. In particular, droplets with linear size $<$($2\pi/k_d$) decay in the presence of coupling with the velocity field; we expect, therefore, that, in the presence of fluid turbulence, the peak of $P(\phi)$ broadens and shifts as it does in our DNS. For a quantitative description of this broadening and the shift of the peak, we must, of course, carry out a full DNS of the Cahn-Hilliard-Navier-Stokes equation as we have done here.

We have investigated, so far, the effect of fluid turbulence on the phase-field
$\phi$ and its statistical properties such as those embodied in $S(k)$ and
$P(\phi)$. We show next how the turbulence of the fluid is modified by $\phi$,
which is an \textit{active} scalar insofar as it affects the velocity field.
In the statistically steady state of our driven, dissipative system, the energy
injection must be balanced by both viscous dissipation and dissipation that
arises because of the interface, i.e., we must have
$\epsilon_{inj}=\epsilon_\nu+\epsilon_{\mu}$.  

\begin{figure*}
\includegraphics[width=0.32\linewidth]{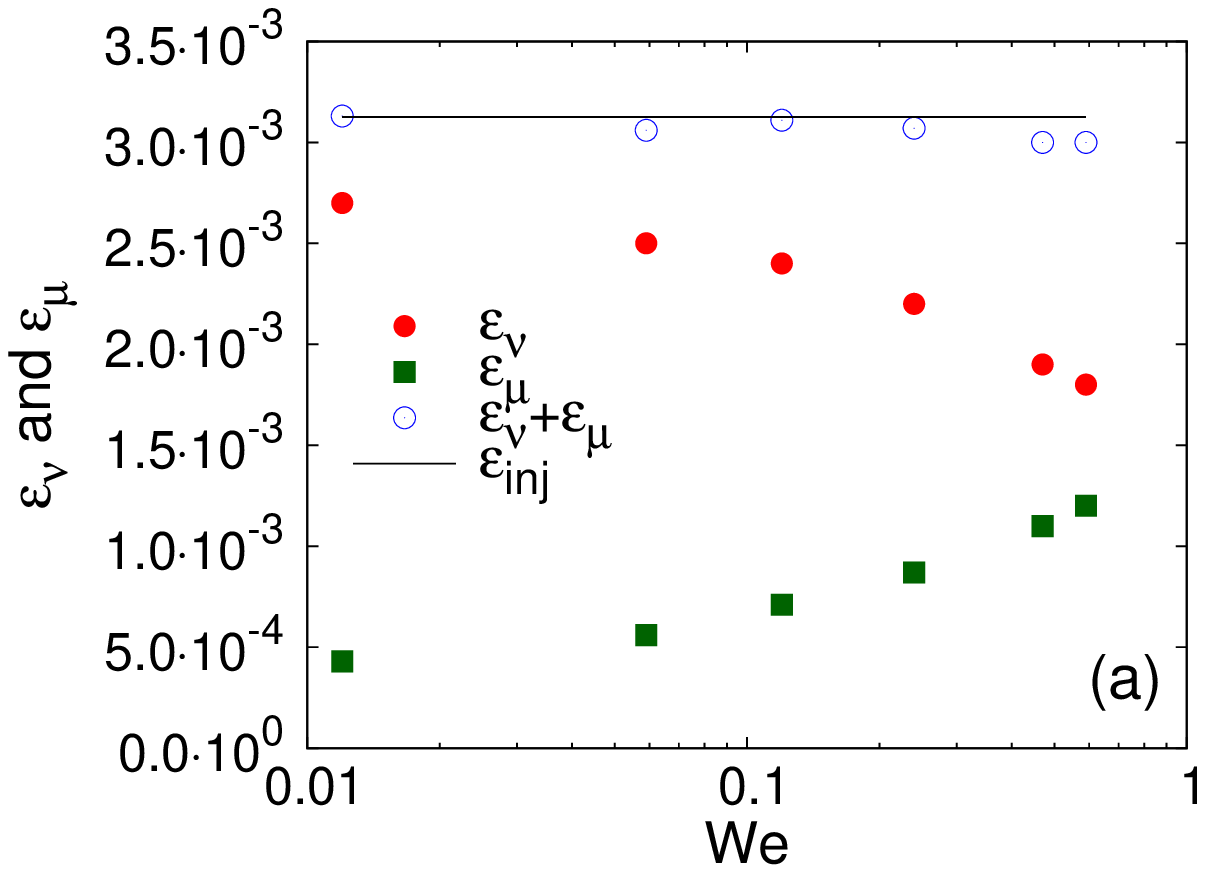}
\includegraphics[width=0.32\linewidth]{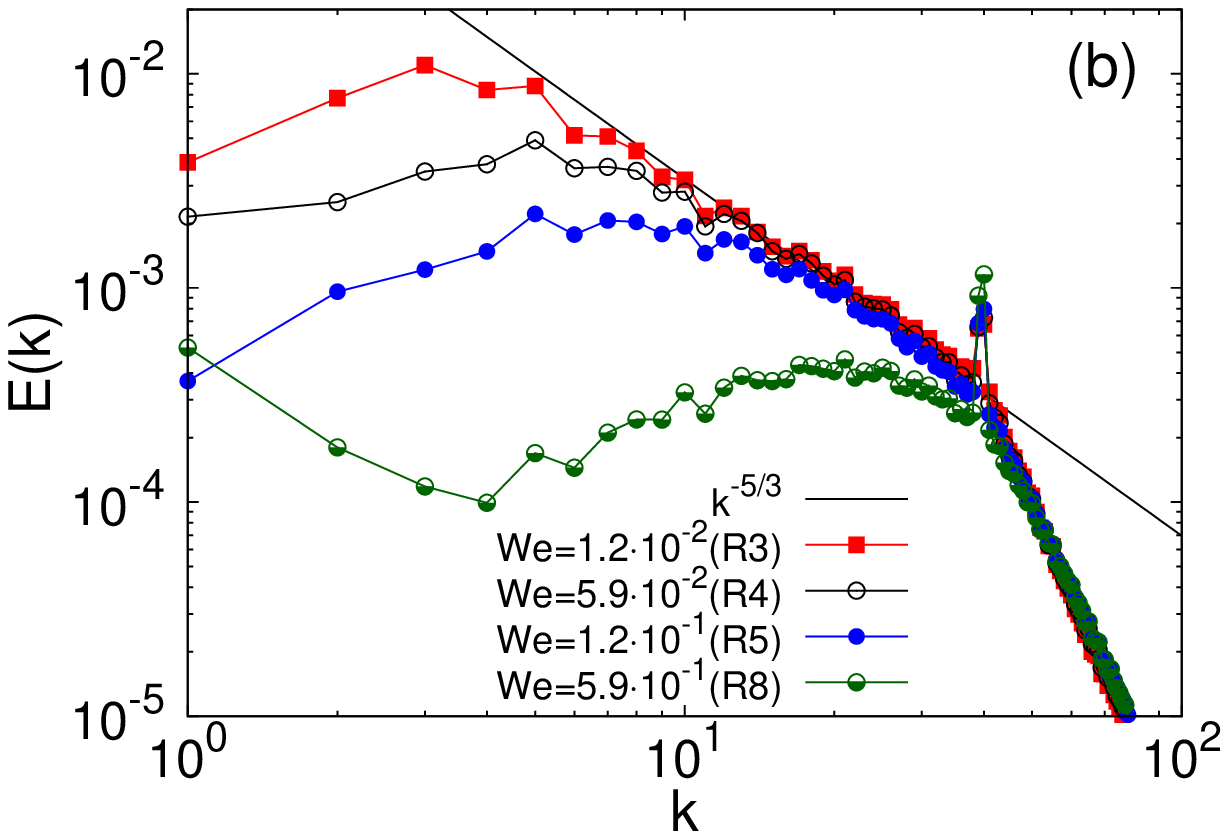}
\includegraphics[width=0.32\linewidth]{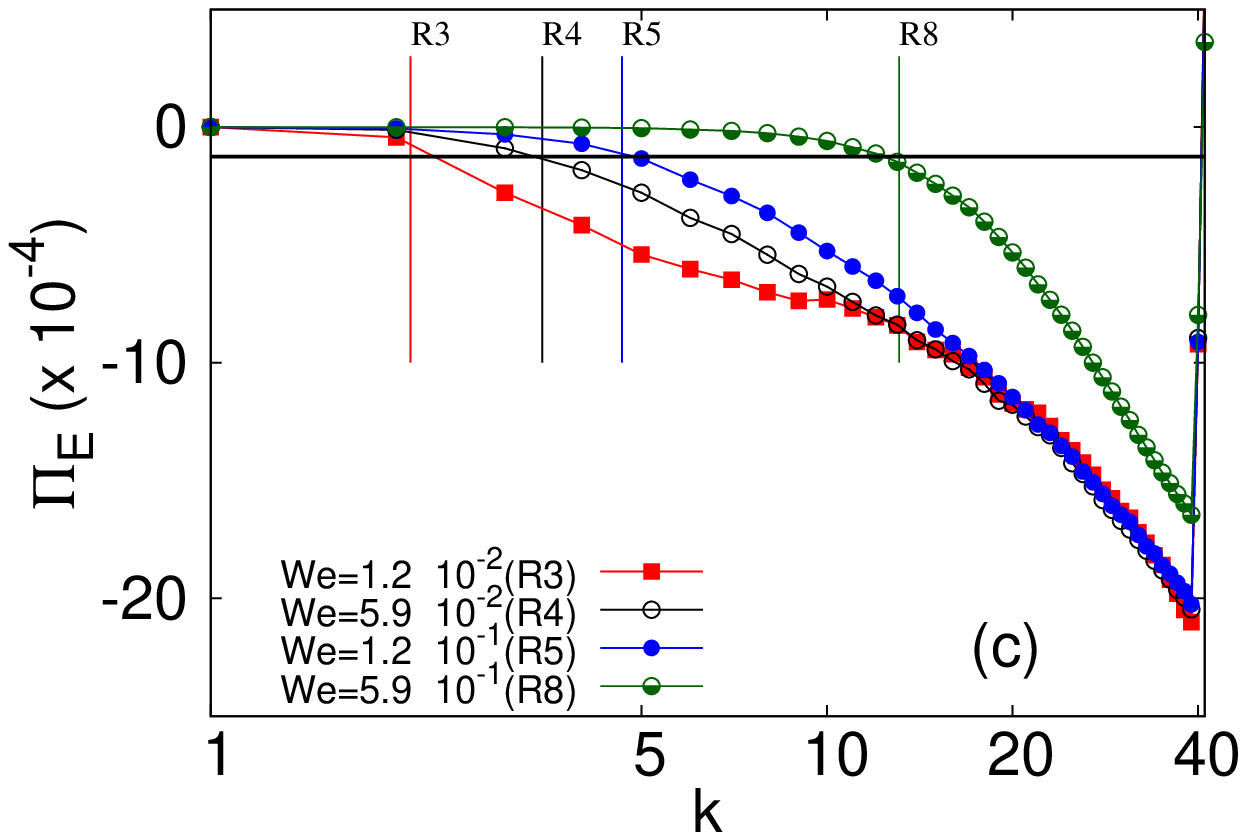}
\caption{\label{fig:spec}(Color online) (a) Plots of the statistically-steady-state values of $\epsilon_{\nu}$, $\epsilon_\mu$, and their sum $\epsilon_{\nu}+\epsilon_{\mu} \simeq \epsilon_{inj}$ versus $We$. (b) 
Log-log (base $10$) plots of the energy spectrum $E(k)$ versus $k$, for different values of $We$, illustrating the truncation of the inverse energy cascade as $We$ increases. The black line indicates the $k^{-5/3}$ for the inverse-cascade in 2D fluid turbulence. (c) Log-log (base $10$) plots of the energy flux $\Pi_E(k)$ versus $k$ for different values of 
$We$. The intersection of the line $0.06\epsilon_{inj}$ (black line) with $\Pi_E(k)$ gives $k_c$, the wave-number at which the inverse energy cascade gets truncated; our estimate of arrest scale $2\pi/L_c$ (vertical lines) is comparable to $k_c$.}
\end{figure*}

In Fig.~\ref{fig:spec}(a), we show that $\epsilon_\nu$ decreases and
$\epsilon_\mu$ increases as we increase $We$, while  keeping $\epsilon_{inj}$
constant, because $L_c$ diminishes (Fig.~\ref{fig:fig1}) and, therefore, the
interfacial length and $\epsilon_\mu$ increase.  This decrease of $L_c$ is
mirrored strikingly in plots of the fluid-kinetic-energy spectrum $E(k)$
(Fig.~\ref{fig:spec}(b)), which demonstrate that the inverse cascade of energy
is effectively blocked at a wavenumber $k_c$, which we determine below, from
the energy flux, and which we find is $\simeq 2\pi/L_c$, where $L_c$ follows
from $S(k)$ (see Fig.~\ref{fig:figh}). The value of $k_c$ increases with $We$;
and the inverse cascade is completely blocked for the largest $We$ we use, for
which $k_c \simeq k_{inj}$, the forcing scale. 

To provide clear evidence that the blocking of the energy flux is closely related to 
the arrest scale, we show in Fig.~\ref{fig:spec}(c) plots of the energy flux 
$\Pi_E(k)=\int_k^\infty T(k^\prime) dk^\prime$ for different values of $We$.
Here $T(k) = \sum_{k-\frac{1}{2} \leq k' \leq k+\frac{1}{2}}\langle\hat{\bm u}(-{\mathbf k},t)\cdot {\bf P}({\bf k}) \cdot \widehat{({\bm u} \times {\bm \omega})({\bf k},t)}\rangle_{t}$ is the energy transfer and ${\bf P}({\bf k})$ is the transverse projector with components $P_{ij}(k)\equiv \delta_{ij}-k_ik_j/k^2$. 
We define $k_c$ as the wave-number at which $\Pi_E(k)$ becomes $4\%$ of  $\epsilon_{inj}$. We find that the wave-numer 
corresponding to the arrest scale $2\pi/L_c$ (marked by vertical lines for each run) is comparable   
to $k_c$. 

\begin{figure}
\includegraphics[width=0.48\linewidth]{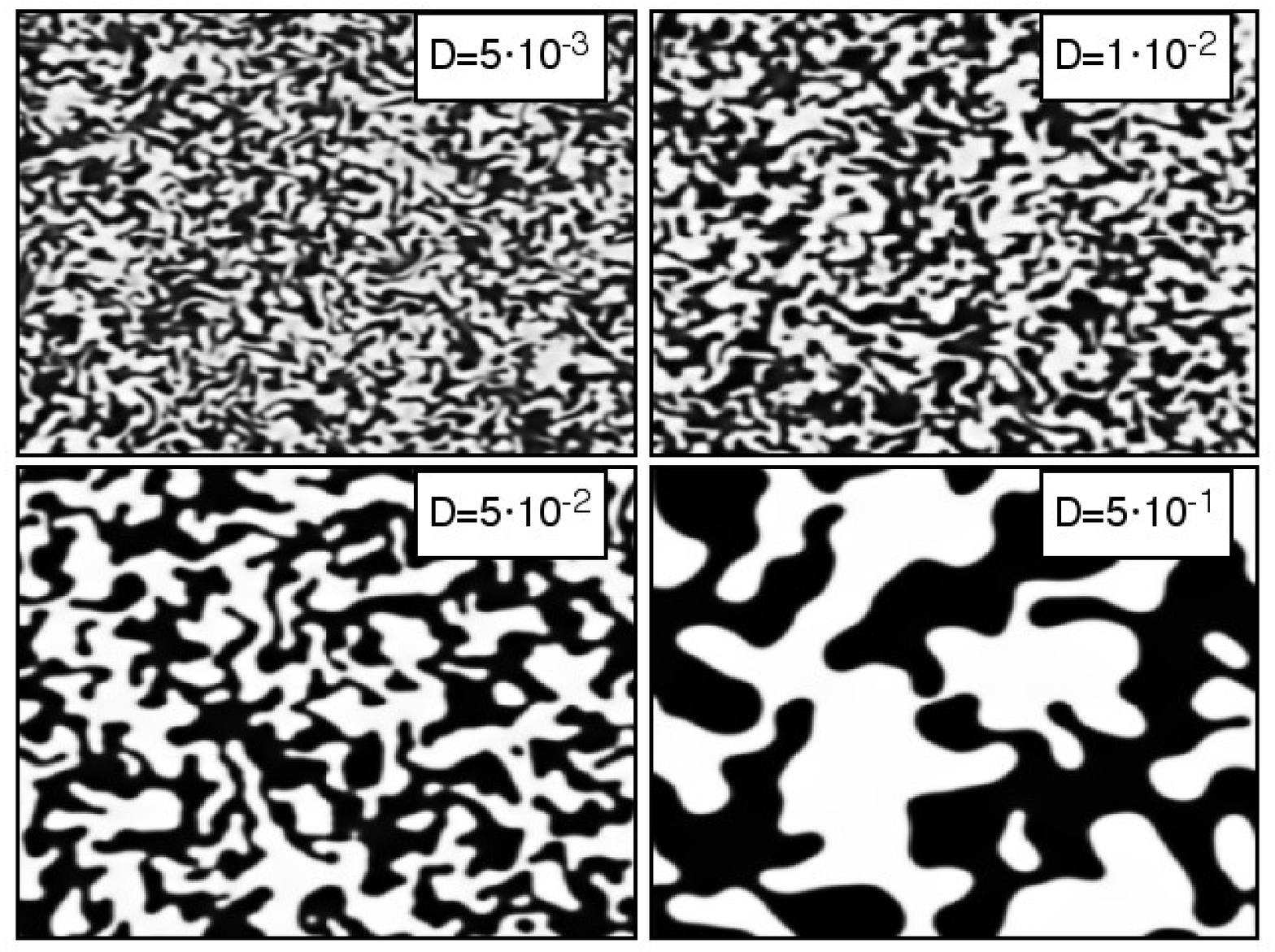}
\includegraphics[width=0.46\linewidth]{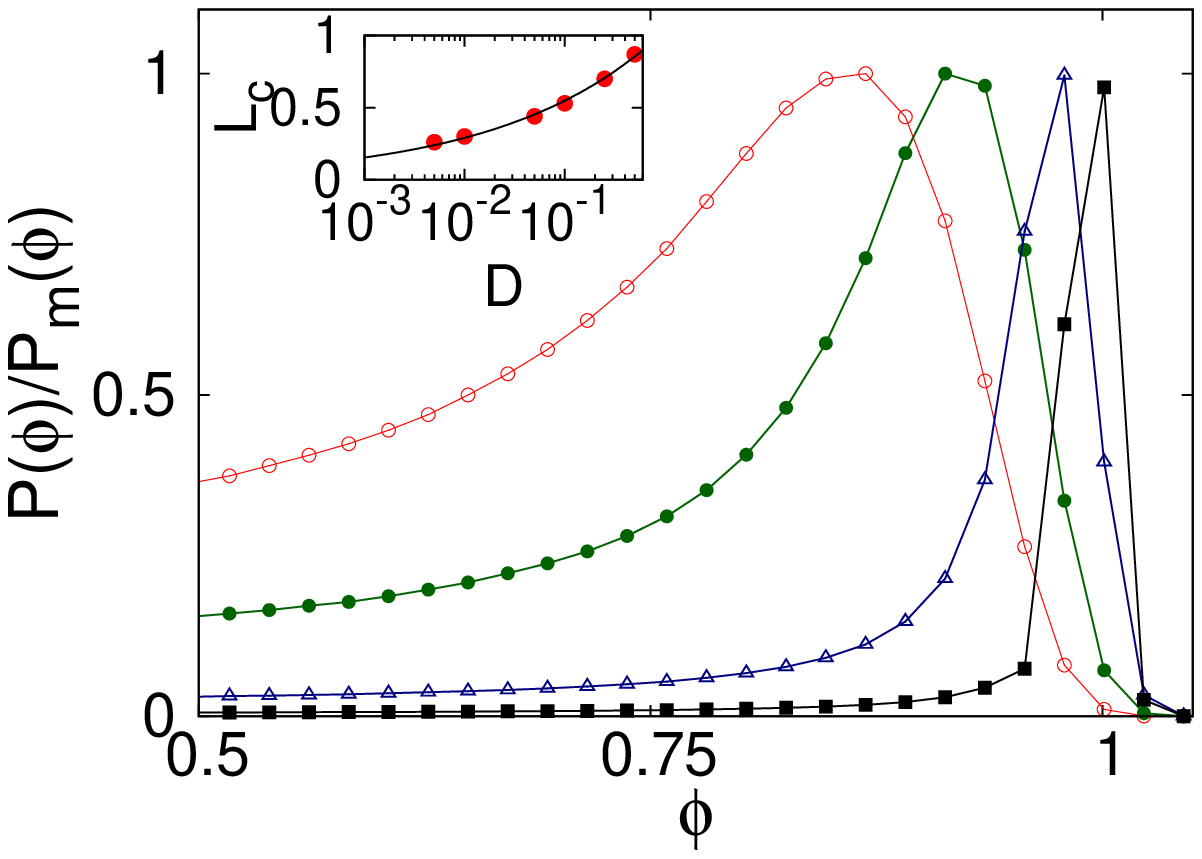}
\caption{\label{fig:pas}(Color online) Passive-advection model: (Left panel) Pseudo-gray-scale plots of the order parameter $\phi$ for different values of diffusivity $D$ (cf.Fig.~\ref{fig:fig1}). (Right panel) Plots of $P(\phi)/P_m(\phi)$, in the vicinity of the maximum at $\phi_{+}$ (cf.Fig.~\ref{fig:figh}(c));the inset shows that $L_c \approx D^{0.27}$ (black line), which is in stark contrast to the Cahn-Hilliard-Navier-Stokes result in the inset of Fig.~\ref{fig:figh}(a).}
\end{figure}

It has been suggested~\cite{ber01,nar07} that coarsening arrest can be
studied by using a model in which the field $\phi$ is advected passively by the
fluid velocity. Such a passive-advection model is clearly inadequate because it
cannot lead to the phase-field-induced modifications in the statistical
properties of the turbulent fluid (see  Fig.~\ref{fig:spec}). However, for the
sake of completeness, we now study the passive-advection case in which the
coupling term $\phi \nabla \mu$ is turned off in Eq.~\eqref{ch:eq2}. We then
contrast the results for this case with the ones we have presented above. The
parameters we use for the passive-advection DNS are $N=1024, \Lambda=\xi^2,
\xi=0.0176$; and we carry out runs for $D=5\cdot10^{-3}, 1\cdot 10^{-2},
5\cdot10^{-2}$ and $5\cdot10^{-1}$. The evolution of the pseudo-grayscale plots of
$\phi$ with $D$, in the left panel of Fig.~\ref{fig:pas}, is qualitatively
similar to the evolution shown in Fig.~\ref{fig:fig1}. There is also a
qualitative similarity in the dependence on $D$ of the scaled PDFs
$P(\phi)/P_m(\phi)$; we can see this by comparing the passive-advection result,
shown in the middle panel of Fig.~\ref{fig:pas} for positive values of $\phi$
in the vicinity of the peak, with its counterpart in Fig.~\ref{fig:figh} (c).
However, there is a qualitative difference in the 
dependence of $L_c$ on $D$: in the passive-advection case we find 
$L_c\sim D^{0.27}$ [Fig.~\ref{fig:pas} (inset)], which is in stark
contrast to the essentially $D$-independent behavior of $L_c$ shown 
in the inset of Fig.~\ref{fig:figh}(c). 

In conclusion, our extensive study of two-dimensional (2D) binary-fluid
turbulence shows how the Cahn-Hilliard-Navier-Stokes coupling leads to an
arrest of phase separation at a length scale $L_c$, which follows from $S(k)$.
We demonstrate that $L_c \sim L_H$, the Hinze scale that we find by balancing
inertial and interfacial-tension forces, and that $L_c$ is independent, within
error bars, of the diffusivity $D$.  We also elucidate how the coupling between the
Cahn-Hilliard and Navier-Stokes equations modifies the properties of fluid
turbulence in 2D. In particular, we show that there is a blocking of the inverse
energy cascade at a wavenumber $k_c$, which we show is $\simeq 2\pi/L_c$.

Earlier DNSs of turbulence-induced coarsening arrest in binary-fluid phase
separation have concentrated on regimes in which there is a forward cascade of
energy in 3D (see Ref.~\cite{per14}) and a forward cascade of enstrophy in 2D
(see Ref.~\cite{ber05}).  Although studies that use a passive-advection model
for $\phi$ obtain results that are qualitatively similar to those we obtain for $S(k)$ and 
the spatiotemporal evolution of $\phi$, they cannot capture the phase-field-induced
modification of the statistical properties of fluid turbulence and the correct
dependence of $L_c$ on $D$. We find our results to be in qualitative agreement with the earlier
studies on advection of binary-fluid mixtures with synthetic chaotic flows
\cite{nar07}; of course, such studies cannot address the effect of the phase field
on the turbulence in the binary fluid.

Some groups have also studied the statistical properties of turbulent, symmetric, binary-fluid 
mixtures above the consolute point, where the two fluids mix even in absence of turbulence~\cite{rui81,jen98}.
In these studies, there is, of course, neither coarsening nor coarsening arrest.

We hope our study will lead to new experimental studies of turbulence in
binary-fluid mixtures, especially in 2D~\cite{muz91,sol96},
to test the specific predictions we make for $L_c$ and the blocking of the
inverse cascade of energy.

We thank S.S. Ray for discussions, the Department of Atomic Energy, the Department of Science and Technology, 
Council for Scientific and Industrial Research, and the University Grants Commision (India) for support.

\bibliographystyle{prsty}

\end{document}

\title{Supplementary material for ``Two-dimensional Turbulence in Symmetric Binary-Fluid Mixtures: Coarsening Arrest by the Inverse Cascade"}
\author {Prasad Perlekar$^{1}$, Nairita Pal$^{2}$, and Rahul Pandit$^{2,3}$} 
\affiliation{${}^1$ TIFR Centre for Interdisciplinary Sciences, 21 Brundavan Colony, Narsingi, Hyderabad 500075, India \\
${}^2$ Centre for Condensed Matter Theory, Indian Institute of Science, Bangalore 560012, India \\
 ${}^3$ Jawaharlal Nehru Centre for Advanced Scientific Research, Jakkur, Bangalore, India}

\maketitle

The parameters for our direct numerical simulations (DNSs) are given in Table~\ref{table1}.
%%%%%%%%%%%%%%%% TABLE OF PARAMETERS  %%%%%%%%%%%%%%%%%
\begin{table}[h]
   \begin{tabular}{@{\extracolsep{\fill}}| c |c | c | c | c | c |c | c |c |c |c |c |c|}
    \hline
    $ $ &$N$ & $\nu$ & $M$  & $\xi (\times 10^{-2})$ & $\Lambda (\times \xi^2)$ & $D$ & $\langle f_\omega \omega\rangle$ &$E$ & $\epsilon_\nu$  
& $\epsilon_{\mu}$  & $We$ & $L_c$  \\
   \hline \hline
%    {\tt NS1}  & $1024$ & $10^{-4}$ & $-$ & $-$ & $-$ &  $5.0$  &  $xx$ & $xx$ & $xx$ & $xx$\\
%    {\tt NS2}  & $1024$ & $10^{-2}$ & $-$ & $-$ & $-$ &  $5.0$ & $xx$ & $xx$  & $xx$ & $xx$\\
    {\tt R1}  & $1024$ & $10^{-4}$ & $10^{-2}$           &  $1.76$ & $1.0$ & $10^{-2}$ &$5.0$ & $3.3\cdot 10^{-2}$ & $2.5\cdot10^{-3}$ & $5.7\cdot10^{-4}$ &  $5.9 \cdot10^{-2}$ & $1.87$ \\
    {\tt R2}  & $1024$ & $10^{-4}$ & $10^{-4}$           &  $1.76$ & $1.0$ & $10^{-4}$&$5.0$ & $3.1\cdot 10^{-2}$ & $2.6\cdot10^{-3}$ & $8.1\cdot10^{-4}$ & $5.9 \cdot 10^{-2}$ & $1.87$\\
    {\tt R3} & $2048$ & $10^{-4}$  & $2\cdot 10^{-4}$& $1.76$  & $5.0$ & $10^{-3}$& $5.0$ & $4.8\cdot 10^{-2}$ & $2.7\cdot 10^{-3}$ & $4.3\cdot10^{-4}$ & $1.2\cdot 10^{-2} $ & $2.97$\\
    {\tt R4} & $2048$ & $10^{-4}$  & $1\cdot10^{-3}$ & $1.76$  & $1.0$ & $10^{-3}$&$5.0$ & $3.0\cdot 10^{-2}$ & $2.5\cdot10^{-3}$ & $5.6\cdot10^{-4}$ & $5.9 \cdot 10^{-2}$ & $1.82$\\
    {\tt R5} & $2048$ & $10^{-4}$  & $2\cdot10^{-3}$ & $1.76$  & $5.0\cdot 10^{-1}$ & $10^{-3}$&$5.0$ & $2.3\cdot 10^{-2}$ &$2.4\cdot10^{-3}$ & $7.1\cdot10^{-4}$  & $1.2\cdot 10^{-1}$ & $1.35$\\
    {\tt R6} & $1024$ & $10^{-4}$  & $4\cdot10^{-3}$ & $1.76$  & $2.5\cdot10^{-1}$& $10^{-3}$& $5.0$ & $1.5\cdot 10^{-2}$ & $2.2\cdot10^{-3}$ & $8.7\cdot10^{-4}$  & $2.4\cdot 10^{-1}$ & $0.9$\\
    {\tt R7} & $1024$ & $10^{-4}$  & $8\cdot10^{-3}$ & $1.76$  & $1.25\cdot 10^{-1}$& $10^{-3}$ & $5.0$ & $9.5\cdot 10^{-3}$ & $1.9\cdot10^{-3}$& $1.1\cdot10^{-3}$  & $4.7\cdot 10^{-1}$ & $0.57$\\
    {\tt R8} & $2048$ & $10^{-4}$  & $10^{-2}$           & $1.76$  & $1.0\cdot 10^{-1}$ & $10^{-3}$ &$5.0$ & $8.1\cdot 10^{-3}$ &$1.8\cdot10^{-3}$ & $1.2\cdot10^{-3}$ & $5.9\cdot 10^{-1}$ & $0.48$\\
    {\tt R9} & $2048$ & $10^{-4}$  & $2\cdot 10^{-4}$& $0.50$ & $5.0$ & $10^{-3}$ &$5.0$ & $3.7\cdot 10^{-2}$ & $2.8\cdot10^{-3}$& $3.4\cdot10^{-4}$  & $4.1\cdot 10^{-2}$ &$1.55$\\
    {\tt R10} & $2048$ & $10^{-4}$  & $10^{-3}$         & $0.50$ & $1.0$ & $10^{-3}$ & $5.0$ & $1.7\cdot 10^{-2}$ & $2.6\cdot10^{-3}$ &$5.3\cdot10^{-4}$  & $2.1\cdot10^{-1}$ &$0.63$\\
  %  {\tt R11} & $1024$ & $10^{-4}$ & $2\cdot10^{-2}$ &  $1.76$ & $5.0\cdot10^{-2}$ & $10^{-3}$ &$5.0$ & $5.3 \cdot 10^{-3}$ & $1.4\cdot10^{-3}$ & $1.6\cdot10^{-3}$ & $0.28$ \\
  % {\tt R12} & $1024$ & $10^{-4}$ & $2.7\cdot10^{-2}$& $1.76$ & $3.75\cdot10^{-2}$ & $10^{-3}$  & $5.0$ & $1.6\cdot 10^{-2}$ & $1.6\cdot 10^{-3}$ & $1.5\cdot10^{-3}$ & $0.22$ \\
  %{\tt R13} & $1024$ & $10^{-4}$ &$3.2\cdot 10^{-2}$ & $1.76$ & $3.125\cdot10^{-2}$ & $10^{-3}$ & $5.0$ & $9.5\cdot 10^{-2}$ & $1.8\cdot10^{-3}$ & $1.1\cdot10^{-3}$ & $0.20$  \\
    {\tt R11}  & $1024$ & $10^{-2}$ & $2\cdot 10^{-4}$ &  $1.76$ & $5\cdot10^2$ & $10^{-1}$& $4\cdot 10^{5}$ & $2.0 \cdot 10^{1}$&$2.4\cdot 10^2$ &  $1.2\cdot10^1$  & $0.22$ & $2.3$\\
    {\tt R12}  & $1024$ & $10^{-2}$ & $10^{-3}$           &  $1.76$  & $10^2$    & $10^{-1}$     & $4\cdot 10^{5}$  & $8.8$ & $2.0\cdot10^2$ &$4.7\cdot10^1$ & $1.1$ & $0.55$\\
    %{\tt S2B}  & $1024$ & $10^{-2}$ & $2\cdot 10^{-6},2\cdot 10^{-5}$ &  $1.76\cdot 10^{-2}$ \\
    %{\tt S2C}  & $1024$ & $10^{-2}$ & $2\cdot 10^{-7},2\cdot10^{-6}$ &  $1.76\cdot 10^{-2}$ \\
 \hline
\end{tabular}
\caption{\label{table1} \small Parameters $N, \nu, M, \xi, \Lambda, D, \langle
f_\omega \omega \rangle$ for our DNS runs {\tt R1-R12}.   The
forcing wave number is fixed at $k_{inj} \equiv 2\pi/\ell_{inj}=40$,  $N^2$ is
the number of collocation points, $\nu$ is the kinematic viscosity, $M$ is the
mobility, parameter, $\xi$ sets the scale of the interface width, the surface
tension $\sigma=2\sqrt{2}/3 (\Lambda/\xi)$, $\langle f_{\omega} \omega \rangle$
is the enstrophy-injection rate, which is related to the energy-injection rate
[$\epsilon_{inj}=\langle f_{\omega} \omega \rangle/k_{inj}^2$], $E=0.5 \sum_k
|u_k|^2$ is the fluid kinetic energy, $\epsilon_\nu=\nu \sum_k k^2 |u_k|^2$ is
the fluid energy dissipation rate, $\epsilon_\mu= M \sum_k k^2 |\mu_k^2|$  is
the energy dissipation rate because of the phase-field $\phi$, $L_c$  is the
coarsening-arrest length scale [Eq.~(3)]. ${\tt R1-R2}$: $Re=124$
and $Sc=1$; ${\tt R3-R10}$: $Re=124$ and $Sc=0.1$; ${\tt R11-R12}$:
$Re=53$ and $Sc=0.1$. In some of our runs we also include a friction term $-\alpha \omega$ on the right-hand-side of Eq.~(1); $\alpha=0.001$ for runs ${\tt R4-R8}$ and zero otherwise.  The Cahn number $Ch=\xi/L=\xi/2\pi$.} 
\end{table}

\bibliographystyle{prsty}
\bibliography{spinodal_ref}